\documentclass[final,5p,times,twocolumn]{elsarticle}

\usepackage{amsmath, amssymb}
\usepackage{csquotes}
\usepackage[switch]{lineno}
\usepackage[colorlinks=true,linkcolor=black,citecolor=black,urlcolor=blue]{hyperref}

\newcommand{\Gr}{\mathrm{Gr}}

\newcommand{\lR}{\lambda_\text{R}}
\newcommand{\muB}{\mu_\text{B}}
\newcommand{\grxy}[2]{$\mathrm{Gr}/\mathrm{#1}/\mathrm{#2}$}
\newcommand{\grall}{$\mathrm{Gr}/\allowbreak(\mathrm{Au},\mathrm{Pt})/\allowbreak(\mathrm{Co},\mathrm{Mn})$}

\journal{Carbon}

\begin{document}
\hypersetup{linkcolor=black,citecolor=black,urlcolor=blue}

\begin{frontmatter}

\title{Magneto-spin--orbit graphene: Chern topology and photoinduced Hall response}

\author[spbu]{A.V.~Tarasov\corref{cor1}}
\ead{artem.tarasov@spbu.ru}

\author[spbu]{A.V.~Eryzhenkov}
\affiliation[spbu]{organization={Saint Petersburg State University},
                   city={Saint Petersburg},
                   postcode={198504},
                   country={Russia}}
\cortext[cor1]{Corresponding author}

\begin{abstract}
Magneto-spin-orbit graphene can host Chern phases due to proximity interactions, 
but in realistic heterostructures these depend on substrate chemistry, 
local stacking registry and may be altered by buried-interface reconstruction. 
Investigation of these effects require special efforts to take into account 
the long-range moir\'{e} patterns that often arise for these heterostructures.
We address this question using first-principles calculations and 
effective four-band models for four out-of-plane magnetized \grall{} 
heterostructures in commensurate $(2\times2)$ and moir\'{e} $(9\times9)$ 
supercells where the latter include misfit dislocations 
for the Co-based interfaces. The commensurate cells overestimate charge 
transfer and exaggerate valley asymmetry, 
whereas moir\'{e} averaging strongly reduces both effects while preserving 
sizable spin splittings. The Pt spacer mediates stronger proximity
interactions than Au, while replacing Co by Mn substantially changes the balance
of the effective exchange interactions in a spacer-dependent manner.
Introduction of misfit dislocations into the Co-based 
structures is selective and tunes the Dirac gap in \grxy{Au}{Co} or spin splittings 
in \grxy{Pt}{Co}. Graphene in \grxy{Pt}{Co} exhibits effective ferromagnetic 
Chern phase with $C = 2$ in all studied supercells.
By contrast, the moir\'{e} models of the other three compositions are ferrimagnetic and topologically trivial, with $C=0$. Nevertheless, they exhibit helicity-dependent photoinduced Hall responses, which are strongly enhanced by misfit-dislocation reconstruction in \grxy{Au}{Co}. 
This work demonstrates
that moir\'{e}-scale stacking controls the effective Chern phase and Hall conductivity, whereas buried-interface defects tune the photoinduced Hall response without changing the topology.
\end{abstract}


\end{frontmatter}


\section{Introduction}

Pristine graphene has a unique gapless, spin-degenerate Dirac spectrum that represents a simple and versatile platform for subsequent manipulation of spin, valley and sublattice degrees of freedom. 
Engineering of proximity effects using various substrates without direct chemical modification of graphene is a promising route to perform those manipulations. The substrate geometry can induce sublattice asymmetry whereas exchange interaction and spin-orbit coupling (SOC) originating from suitable substrates can be induced in graphene through proximity effects \cite{rashba2009,rybkin2018,eryzhenkov2023}.
As a result, a diverse range of topological or valley-selective phases in graphene can be implemented
\cite{eryzhenkov2023,haldane1988,phong2018,takenaka2019,zhou2021}. 

Several strategies have been explored to induce magnetism in graphene.
Direct contact with Co or Ni produces strong $\pi$--$d$ hybridization that leads to significant spin polarization of graphene states but also obscures or even destroys the Dirac cone
\cite{rader2009,varykhalov2009,dedkov2010,marchenko2015,usachov2015}. Magnetic insulating substrates such as 
EuO offer a less invasive alternative that largely preserves the Dirac cone integrity \cite{swartz2012,averyanov2018},
but the resulting band structure becomes very sensitive to details of 
the atomic structure of the system \cite{yang2013,hallal2017}.
Magnetism can also be created by vacancies or chemisorbed light atoms
\cite{yazyev2007,cervenka2009,ugeda2010,nair2012,boukhvalov2008,hong2012,
giesbers2013,gonzalez2016}. These approaches demonstrate atomic-scale control of local magnetic moments, but a spatially uniform magnetic state requires precise control over defect formation.

Addition of a heavy metal spacer layer between graphene and a ferromagnet is attractive because it can simultaneously transmit exchange interaction and induce SOC in graphene while protecting its Dirac cone from chemisorption. Au intercalation restores quasifreestanding character of graphene on Ni and Co, allowing its doping and spin
structure to be tuned
\cite{varykhalov2008,sanchezbarriga2010,varykhalov2010} and induces large Rashba-like spin splittings in graphene
\cite{marchenko2012,shikin2013,varykhalov2015}. 
Similar SOC enhancement has also been obtained by proximity to Pb- and Pt-containing interfaces
\cite{calleja2015,klimovskikh2017}. These experiments established
heavy-metal intercalation as a powerful route to obtain graphene with significant external SOC, which is important because its intrinsic SOC is only of order tens of $\mu$eV \cite{konschuh2010}.

First-principles studies have shown that apparent spin splitting of the Dirac cone is tied to $\pi$--$d$ hybridization (rendered by numerous avoided crossings between the $\pi$-bands and metallic spacer $d$-bands) which strongly depends on lateral alignment of graphene and the spacer layer (local stacking registry) \cite{marchenko2012,slawinska2018,voloshina2018,slawinska2019,krivenkov2017,lopez2019}. 
Thus, the atomic SOC of the spacer alone does not determine the induced spin splitting or the low-energy topology, which also depend on interface-specific hybridization and local stacking.
These findings are important for \emph{ab initio}
material modeling of such compound substrates that can be performed 
using small \enquote{commensurate} supercells or more realistic large moir\'{e} supercells. The first approach provides a coarse-grained
overview of system properties that depend on its chemical composition
but samples only a small set of perfect local stacking registries and misses the moir\'{e} structure that effectively averages over them
and folds the Brillouin zone (BZ) further
\cite{rybkin2018,marchenko2012,slawinska2018,voloshina2018,usachov2014}.
The case of \grxy{Au}{Co}, for which STM and LEED experiments establish a reconstructed large-period interface \cite{rybkin2018,rybkin2022},
highlights the importance of large supercells for capturing fine details of band and spin structures of such heterostructures which we refer to as \emph{magneto-spin-orbit graphene}. Their defining feature is a quasifreestanding Dirac cone with a complex spin structure arising from proximity-induced exchange and spin-orbit interactions.

Magneto-spin-orbit graphene heterostructures can generally realize different magnetization configurations depending on the substrate and interface structure. For example, the \grxy{Au}{Co} system studied in Ref.~\cite{rybkin2022} exhibited an in-plane magnetization of the Co film. Subsequent STM and STS measurements on the same system under an out-of-plane magnetic field of 1~T showed that the sublattice contrast and a Dirac gap of approximately 30~meV persist when the Co magnetization is oriented perpendicular to the graphene plane~\cite{rybkin2025nanoscale}. Such an out-of-plane magnetic configuration is of particular interest because it can enable graphene phases with a nonzero Chern number~\cite{eryzhenkov2023}. Moreover, the magnitude of the Chern number gives the net number of chiral edge channels: phases with $|C|>1$ can therefore support multichannel chiral transport. This has motivated extensive theoretical and experimental efforts to realize high-Chern-number phases, including in magnetic topological-insulator multilayers and multilayer graphene~\cite{Wang2013HigherPlateaus,Zhao2020TuningChern,Han2024LargeQAH,Wang2026HighChern}. In ultrathin Co films, perpendicular magnetization can in principle be achieved, although the resulting magnetic state is sensitive to the film thickness, the choice of heavy-metal spacer, stacking faults, and even the specific synthesis procedure used to prepare the system~\cite{rougemaille2012,yang2016pma,vu2016,blancorey2021,gogina2026}.

In this work, we systematically compare four heterostructures \grall{} 
with out-of-plane magnetization to assess how their band structures depend on substrate chemistry, formation of large moir\'{e} superstructures and presence of misfit dislocations in the Co-based interfaces. The \grxy{Au}{Co} system with 
an Au spacer and a Co ferromagnet serves as a base system. 
The choice of a Pt spacer instead of Au may allow for stronger SOC induction on graphene
\cite{marchenko2012,shikin2013,varykhalov2015,slawinska2018,klimovskikh2014},
whereas a Mn ferromagnetic underlayer has a substantially higher magnetic moment
than that of Co. It is understood that the proximity exchange between 
graphene and transition metal depends on details of the corresponding 
$\pi$-$d$ hybridization \cite{rader2009,marchenko2015}; therefore,
substitution of Co with Mn may shed light on how these factors 
compete with the larger Mn magnetic moment.

We first consider small $(2 \times 2)$ commensurate supercells to isolate the coarse-grained effects of substrate
composition. Second, large $(9 \times 9)$ moir\'{e} supercells are used
to investigate how the more realistic moir\'{e} representation reshapes the 
electronic structure of graphene. Finally, these \emph{ab initio} band structures of graphene are fitted using
an effective $4 \times 4$ tight-binding model of magneto-spin-orbit
graphene, which allows us to analyze effective exchange, SOC and sublattice
polarization and to perform computations of Chern numbers and 
photoinduced Hall responses.

\section{Methods}

\begin{figure}[!htbp]
  \centering
  \includegraphics[width=\columnwidth]{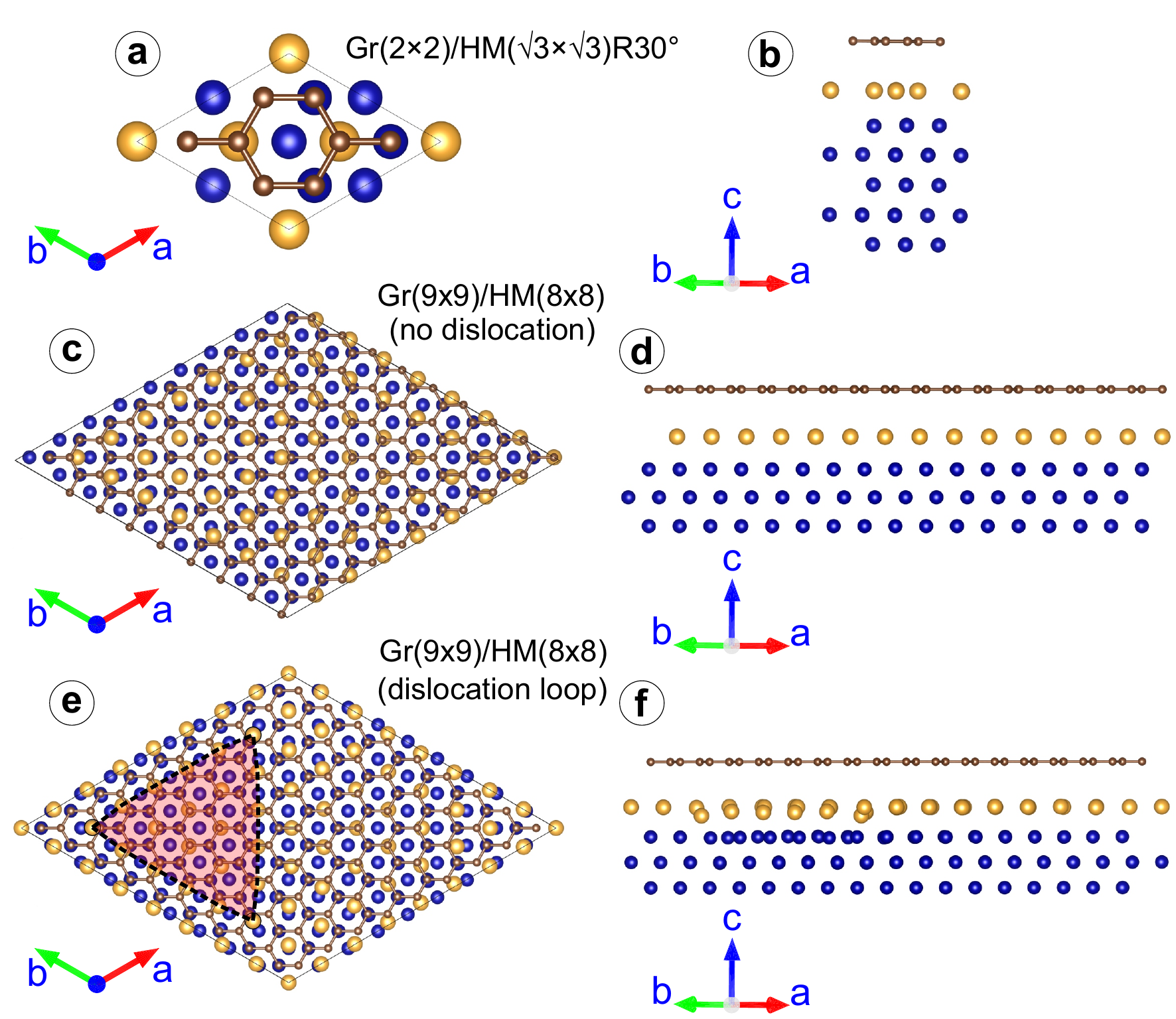}
  \caption{Top and side views of common structural 
  models used for all studied \grxy{HM}{FM} heterostructures where HM (yellow) 
  is a spacer heavy metal (Au,~Pt) and
  FM (blue) is a ferromagnet (Co,~Mn). 
  Panels (a,~b) show the corresponding views
  of commensurate 
  $\mathrm{Gr}(2\times2)/\mathrm{HM}(\sqrt3\times\sqrt3)R30^\circ$ cell,
  panels (c,~d) show the moir\'{e} supercell 
  $\mathrm{Gr}(9\times9)/\mathrm{HM}(8\times8)$
  and panels (e,~f) show the moir\'{e} supercell used only with 
  $\text{FM} = \text{Co}$ 
  where the triangle marks a misfit dislocation 
  similar to that found in \grxy{Au}{Co} according to STM 
  data in Ref.~\cite{rybkin2022}.}
  \label{fig:structures}
\end{figure}

\subsection{Structural models}

Two cell sizes were used for all four systems \grall{}, while the
reconstructed $9\times9$ model was considered only for the Co-based
systems, as summarized in Fig.~\ref{fig:structures}. In the
commensurate $(2 \times 2)$ model in Figs.~\ref{fig:structures}(a,~b)
a $(2 \times 2)$ graphene supercell is matched to a
$(\sqrt3\times\sqrt3)R30^\circ$ heavy metal spacer as in
Ref.~\cite{rybkin2018}. There are three inequivalent stacking
registries: one spacer atom lies below the center of a graphene hexagon
whereas the other two spacer atoms each lie below carbon atoms of $A$
and $B$ sublattices. The same stacking was used for all four
compositions. The commensurate cell therefore serves as a controlled
fixed stacking pattern for the chemical comparison rather than as a representation of
the complete moir\'{e} structure.

In the large moir\'{e} supercells in Figs.~\ref{fig:structures}(c--f)
the $9\times9$ graphene supercell is matched to an $8\times8$
heavy metal spacer. This choice represents the large periodicity
observed by STM and LEED for \grxy{Au}{Co}
\cite{rybkin2018,rybkin2022}.
For the chemical comparison, the defect-free \grxy{Au}{Co}
geometry was used as a structural template for all four compositions:
Au was replaced by Pt and Co by Mn, separately and in combination,
while the atomic coordinates were retained.
The crystal model was obtained for \grxy{Au}{Co} in
Ref.~\cite{rybkin2022} by direct structural optimization which yielded
the Gr--Au separation of approximately 3.4~\AA, but the calculated
Rashba splitting at this distance was substantially smaller than that
inferred from experiment. This underestimation of proximity SOC is
known to arise in idealized models since it strongly depends on the
graphene--metal separation and local stacking registry which is
well-documented for Gr/Au and similar high-$Z$ interfaces with graphene
\cite{marchenko2012,slawinska2018,voloshina2018,slawinska2019}.
We therefore set a common graphene--spacer distance of 3.3~\AA{}
in all cells by rigidly shifting the graphene sheet while leaving the
metal-layer geometry unchanged.

The supercells with misfit dislocations in
Figs.~\ref{fig:structures}(e,~f) are constructed according to the STM
data in Ref.~\cite{rybkin2022} and reflect the corresponding tendency
of the buried Au/Co and Au/Ni interfaces to reconstruct so as to avoid
configurations with Au directly atop Co or Ni
\cite{rybkin2018,nielsen1995,jacobsen1995}.
Because a similar reconstruction may also occur at the Pt/Co
interface, we additionally consider a modeled Pt-substituted
counterpart of the reconstructed \grxy{Au}{Co} geometry, obtained by
replacing Au with Pt while retaining the atomic coordinates. At fixed
composition, supercell size, and graphene--spacer distance, each
defect-free/reconstructed Co-based pair therefore differs only in the
buried interface geometry.

\subsection{First-principles calculations}
\label{sec:dft-method}

The first-principles density functional theory (DFT) calculations were
performed at the Computing Center of the Research Park of Saint
Petersburg State University using the OpenMX~3.9.9 code. 
OpenMX employs a
linear combination of localized
pseudoatomic orbitals \cite{ozaki2003variationally,ozaki2004numerical,
ozaki2005efficient}. We used the GGA-PBE exchange--correlation functional
\cite{perdew1996generalized} together with fully relativistic
norm-conserving pseudopotentials \cite{troullier1991efficient}
allowing for full SOC treatment.
The initial out-of-plane magnetization was used for
all studied systems, but a 0.3~eV magnetic penalty was applied specifically
to the carbon spins in \grxy{Au}{Mn} to enforce their strictly out-of-plane
orientation.

The element basis sets were specified as
C6.0-s2p2d1, Co6.0H-s3p2d1, Au7.0-s3p2d2f1, Mn6.0-s3p2d1, 
Pt7.0-s3p2d2f1, where the number gives the orbital 
cutoff radius in Bohr radii and the
subsequent description specifies the number of 
radial functions for each angular
momentum channel. The \enquote{H} label in the Co basis set
specification means that the semicore $3s$ shell is 
treated as a valence shell. 
The localized Mn~$3d$ states were treated within the
Dudarev scheme of DFT+$U$ formalism 
\cite{han2006ldau,dudarev1998} using $U_{3d} = 5$~eV. 

The Brillouin zone was sampled with a $6\times6\times1$ Monkhorst--Pack grid for the $2\times2$ cells and only at the $\Gamma$ point for the $9\times9$ supercells. 
The real-space integration grid was defined by a cutoff of $200$~Ry
and the self-consistent energy convergence criterion was set to $3 \times10^{-5}$~Hartree.

\subsection{Model Hamiltonian of graphene}

DFT band structures of graphene on the studied substrates were
fitted to a $4 \times 4$ tight-binding Hamiltonian. This Hamiltonian was
previously used for $\Gr/\mathrm{Au}/\mathrm{Co}$ in Ref.~\cite{rybkin2022}
and analyzed in detail in Ref.~\cite{eryzhenkov2023}.
Here, it is extended by the spin-independent Semenoff mass term
\cite{semenoff1984} which accounts for sublattice polarization
for more accurate band structure fitting. The model in the 
$\{A\!\uparrow,A\!\downarrow,B\!\uparrow,B\!\downarrow\}$ basis reads
\begin{equation} \label{eq:hamiltonian}
  \mathcal H(\mathbf k) = E_\text{D} -
  \begin{pmatrix}
    \mathcal S_z u_A - \Delta_\text{S} &
    \varepsilon^0_{\mathbf k} e^{-i\vartheta_{\mathbf k}} + \mathcal{R}_{\mathbf k} \\
    \varepsilon^0_{\mathbf k} e^{ i\vartheta_{\mathbf k}} + \mathcal{R}_{\mathbf k}^\dagger &
    \mathcal S_z u_B + \Delta_\text{S}
  \end{pmatrix}.
\end{equation}
Each displayed block is a $2 \times 2$ matrix in spin space. Addition of
scalars to matrices is implied to involve unit matrices of
necessary size. Nontrivial $2 \times 2$ matrices such as the standard
Pauli matrices $\mathcal S_i$ are written in calligraphic letters.
The overall constant term $E_{\mathrm D}$ is used to
align the fitted model spectrum with the DFT band structures.
For the topological and optical calculations, this uniform energy
shift was removed and the internal energy zero was placed inside the
fitted gap.
The pristine-graphene dispersion $\pm \varepsilon^0_{\mathbf k}$
corresponds to conventional nearest-neighbor (NN) hopping of strength
$t$ and is expressed through the graphene structure factor
\begin{equation*}
 f(\mathbf{k}) = |f(\mathbf{k})| e^{-i\vartheta_{\mathbf{k}}} = 
 \sum_{j=1}^{3} e^{-i \mathbf{k} \mathbf{d}_j},
 \quad
 \varepsilon^0_{\mathbf{k}} = t|f(\mathbf{k})|,
\end{equation*}
where $\mathbf{d}_j$ are the three $A \to B$ NN hopping vectors of the
same length $d$. The additional spin-dependent off-diagonal block
accounts for the NN Rashba spin-orbit coupling (SOC) of strength $\lR$
and has the matrix form 
\begin{equation*}
 \mathcal{R}_{\mathbf{k}} = \frac{\lR}{3d}
 \left[ 
 \mathcal S_x \partial_y f(\mathbf k) - 
 \mathcal S_y \partial_x f(\mathbf k) 
 \right], \quad \partial_i \equiv \partial_{k_i}.
\end{equation*}
Finally, the diagonal blocks describe the collinear 
out-of-plane proximity exchange where $u_A$ and $u_B$ are
sublattice exchange energies and their plus and minus signs specify
whether they prefer $+z$ or $-z$ magnetization of the corresponding
sublattice. Following Ref.~\cite{eryzhenkov2023}, the uniform and
staggered exchange components are, respectively,
\begin{equation*}
 u_{\mathrm{FM}} = \frac{u_A+u_B}{2}, \qquad 
 u_{\mathrm{AFM}} = \frac{u_A-u_B}{2}.
\end{equation*}

A separate intrinsic SOC term is not 
included in this work even at the effective level in order to keep 
the discussion focused on the external proximity effects.
The validity of this approach is assessed by direct comparison 
with the DFT band structures. 

DFT band structure fitting was performed by selecting $E_{n\mathbf{k}}$
points in the low-energy region where the corresponding solution
$\psi_{n\mathbf{k}}$ has the largest carbon weight and then
adjusting the model parameters such that the resulting model
band structure reproduces the corresponding graphene-derived
branches. Their assignment was guided by carbon weight, branch
continuity, and spin polarization. The two valleys were fitted
simultaneously for the moir\'{e} structures, whereas separate
valley fits were used for the strongly valley-asymmetric commensurate
Co-based systems, as discussed below.
The fit quality is quantified by the root-mean-square error (RMSE) in
band energies.

All computations of the Chern numbers using the four-band model were
performed on a uniform $1201 \times 1201$ BZ grid using the
nonabelian Fukui--Hatsugai--Suzuki (FHS) method \cite{fukui2005},
with the occupied subspace chosen to consist of the two bands with the lowest energy.
Whenever a Chern number is assigned, the two occupied bands are separated by a finite global gap from the other two bands. The grid size was determined to yield convergent results
which remained constant upon further grid refinement.

\subsection{Photoinduced Hall response}

The velocity-form expression for the Berry curvature of band $n$ is
\begin{equation*}
 \Omega_{n \mathbf k} = \sum_{m \ne n}
 \frac{2 \operatorname{Im} (v_{nm}^{\mathbf k,x} v_{mn}^{\mathbf k,y}) }
 {(E_{n \mathbf k} - E_{m \mathbf k})^2}, \quad
 v^{\mathbf k,i}_{nm} \equiv \langle u_{n\mathbf k}|\partial_i \mathcal H|u_{m \mathbf k}\rangle
\end{equation*}
and the Kubo formula gives the DC Hall conductivity
\begin{equation*}
  \sigma_{xy} = \frac{e^2}{h} \sum_{n}
  \int_\text{BZ} \mathrm{d}^2\mathbf{k} \, f_{n \mathbf k} \frac{\Omega_{n \mathbf k}}{2 \pi},
  \quad h \equiv 2 \pi \hbar.
\end{equation*}
where $f_{n \mathbf k}$ is the $n \mathbf k$-state occupation factor. 

Our graphene tight-binding model has two valence and two conduction
bands and their equilibrium occupations are set to $f^0_{v \mathbf k} = 1$
and $f^0_{c \mathbf k} = 0$. However, some systems are topologically trivial
(i.e., their Chern number $C = 0$) and have $\sigma_{xy} = 0$, although their Berry 
curvature does not vanish identically. In order to detect a transverse 
response caused by that curvature, a light photoexcitation can be used
to populate conduction bands with electrons and valence bands with holes 
transiently to obtain nonequilibrium carriers undergoing 
transverse motion under a longitudinal DC field. 
This approach for obtaining nontrivial response from formally 
topologically trivial systems was used previously in Refs.~\cite{zhou2021,rybkin2022,dai2007,priyadarshi2015,sato2019,murotani2023}.

In this work, we compute a normalized photoexcitation response by the following
procedure. The pulse fluence $\mathcal F$, defined as the pulse energy per unit
sample area, is related to the optical field intensity $I$ through the relation
$\mathrm d\mathcal F/\mathrm dt = I$. Therefore, the fluence-normalized
DC Hall response is
\begin{equation*}
  \frac{\Delta \sigma_{xy}}{\mathcal F} \equiv 
  \left.\frac{\mathrm d \sigma_{xy}}{I \, \mathrm d t}\right|_{t = 0} =
  \frac{e^2}{h} \sum_{n}
  \int_\text{BZ} \mathrm{d}^2\mathbf{k} \, \left. \frac{\mathrm df_{n \mathbf k}}{I \, \mathrm dt} \right|_{t = 0}
  \frac{\Omega_{n \mathbf k}}{2 \pi}.
\end{equation*}
The master equation for state occupations at the excitation instant is
\begin{equation*}
  \left.\frac{\mathrm df_{n \mathbf k}}{\mathrm dt}\right|_{t = 0} = 
  \sum_{m \ne n} \left(W^\mathbf{k}_{m \to n} f^0_{m \mathbf k} - 
  W^\mathbf{k}_{n \to m} f^0_{n \mathbf k}\right)
\end{equation*}
where $W$ denote the vertical transition rates. Therefore,
\begin{equation*}
  \left. \frac{\mathrm df_{c \mathbf k}}{\mathrm dt} \right|_{t = 0} =  \sum_{v} W^\mathbf{k}_{v \to c}, \quad
  \left. \frac{\mathrm df_{v \mathbf k}}{\mathrm dt} \right|_{t = 0} = -\sum_{c} W^\mathbf{k}_{v \to c}.
\end{equation*}
The transition rates are estimated using the Fermi golden rule, assuming 
vertical optical transitions and photon energy $E = \hbar \omega$:
\begin{equation*}
\begin{split}
  W^{\mathbf{k},\pm}_{v \to c} &= 
  2 \pi^2 \alpha I \times \\ &\times \int \mathrm dE \, \frac{g_\sigma(E - E_0)}{E^2}
  |v_{vc}^{\mathbf k,\pm}|^2 \delta(E_{c \mathbf k} - E_{v \mathbf k} - E) = \\
  &= 2 \pi^2 \alpha I 
  \frac{g_\sigma(E_{c \mathbf k} - E_{v \mathbf k} - E_0)}
  {(E_{c \mathbf k} - E_{v \mathbf k})^2} |v_{vc}^{\mathbf k,\pm}|^2
\end{split}
\end{equation*}
where $\alpha$ is the fine structure constant, 
the symbols $p_x\pm ip_y$ denote the two circular-polarization channels,
whose matrix elements are
$v_{mn}^{\mathbf k,\pm} = v_{mn}^{\mathbf k,x} \pm i v_{mn}^{\mathbf k,y}$.
The normalized Gaussian spectral line shape is
\begin{equation*}
  g_\sigma(\varepsilon) \equiv 
  \frac{1}{\sigma \sqrt{2 \pi}} \exp \left(-\frac{\varepsilon^2}{2\sigma^2}\right),
  \qquad \int \mathrm dE \, g_\sigma(E-E_0)=1,
\end{equation*}
where $\sigma$ is the broadening and $E_0$ is the central photon energy;
consequently, $\int \mathrm dE \, I g_\sigma(E-E_0)=I$.
The final expression for the fluence-normalized photoinduced Hall response is
\begin{equation}
\begin{split}
  \frac{\Delta \sigma_{xy}^{\pm}}{\mathcal F} &= 
  \frac{e^2}{h} \sum_{vc} \int_\text{BZ} \mathrm{d}^2\mathbf{k} \,
  (\Omega_{c \mathbf k} - \Omega_{v \mathbf k}) \times \\ 
  &\times
  \frac{\pi \alpha |v_{vc}^{\mathbf k,\pm}|^2}
  {(E_{c \mathbf k} - E_{v \mathbf k})^2} 
  g_\sigma(E_{c \mathbf k} - E_{v \mathbf k} - E_0).
\end{split}
\end{equation}
The factor after $e^2/h$ has the dimensions of $\text{\AA}^2 \text{eV}^{-1}$
and is converted to inverse fluence units using the relation
\begin{equation*}
  1 \, \mu\text{J}\,\text{cm}^{-2} \approx 6.2415 \times 10^{-4} \, \text{eV}\,\text{\AA}^{-2}
\end{equation*}
and the resulting response is reported in units of
$(e^2/h)/(\mu\text{J}\,\text{cm}^{-2})$.

BZ integrals were computed separately (and then summed) over 
two square patches for each valley using a $401 \times 401$ grid. Each patch 
had a width equal to 3\% of the intervalley distance $8 \pi/(3 a)$, where 
$a = d \sqrt 3 \approx 2.4962$~\AA{} is the graphene lattice constant. 
The patch and grid sizes were determined to yield 
convergent results stable under grid refinement or patch enlargement. All spectra reported below were calculated with the normalized
Gaussian profile $\sigma=0.10\hbar\omega$.

\section{Results}

\subsection{Commensurate $2\times2$ structures}

\begin{figure*}[!t]
  \centering
  \includegraphics[width=0.84\textwidth]{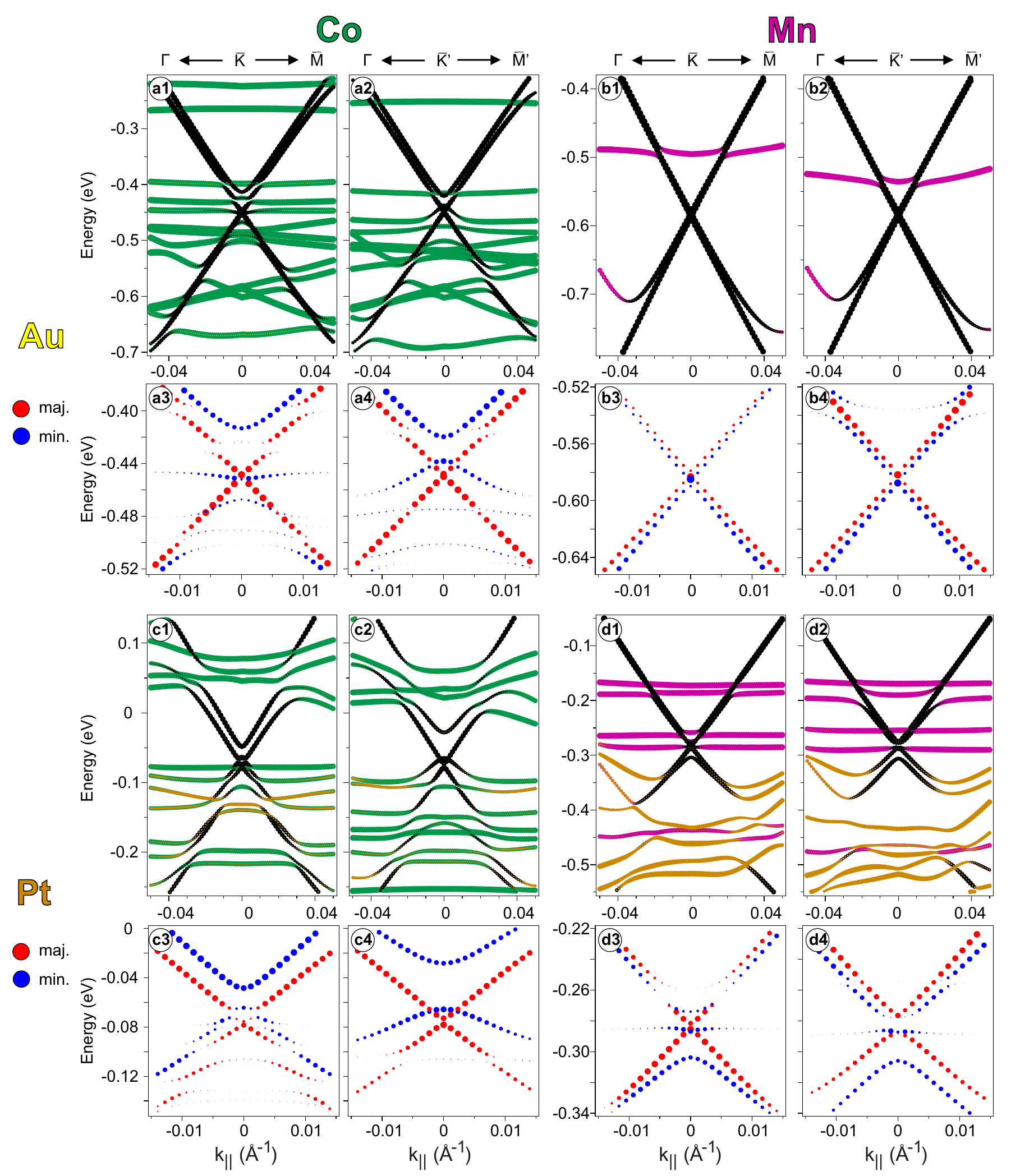}
  \caption{Element- and spin-resolved DFT band structures of the
  commensurate $(2\times2)$ interfaces.  The spacer and magnetic layers are
  identified by the colored row and column headings, respectively.  Within
  each system, panels 1 and 3 correspond to $\overline K$, whereas panels 2
  and 4 correspond to $\overline K'$.  Panels 1 and 2 show the
  element-projected bands; panels 3 and 4 show the corresponding
  carbon-projected spin-resolved bands on enlarged energy and momentum scales.
  Black markers denote carbon, while the metal colors follow the row and
  column labels.  Marker size in panels 1 and 2 gives the element-projected
  weight.  In panels 3 and 4, red (maj.) and blue (min.) denote opposite
  out-of-plane spin polarizations of the carbon states, while marker size
  represents their magnitude.  Energies are relative to the Fermi level.}
  \label{fig:orbital}
\end{figure*}

Figure~\ref{fig:orbital} shows the $(2\times2)$ supercell DFT band structures of all four \grall{} systems in both graphene valleys resolved into elemental contributions. The magnetic moments of Mn and Co atoms in the ferromagnetic layers are approximately $5~\muB$ and $1.6~\muB$, respectively. In all cases, the Dirac cones of graphene are quite recognizable, but their doping, hybridization and low-energy spin structures near the Dirac points differ substantially. Table~\ref{table:dft2x2} summarizes the Dirac point (DP) positions and spin splittings in each valley.

All four Dirac cones are $n$-doped, although the shift is substantially
smaller in \grxy{Pt}{Co} [Figs.~\ref{fig:orbital}(c1) and
\ref{fig:orbital}(c2)].  Within each spacer pair, replacing Co by Mn moves the
DP farther below the Fermi level and accompanies a change in interfacial band
alignment and charge redistribution.
On the other hand, \grxy{Au}{Mn} shows the largest doping
[Figs.~\ref{fig:orbital}(b1) and \ref{fig:orbital}(b2)] together with
the weakest apparent hybridization between graphene and the substrate,
resulting in the cleanest Dirac cone among the four systems.
Its Co-based counterpart, the \grxy{Au}{Co} [Figs.~\ref{fig:orbital}(a1) and \ref{fig:orbital}(a2)] system, features
a significantly larger number of avoided crossing structures because more
Co bands lie near the DP, but the hybridization gaps themselves are 
rather small and the graphene Dirac cone mostly retains its structure.
This survival contrasts with direct graphene contact with Co(0001), where
strong $\pi$--$d$ hybridization produces a strongly reconstructed single-spin
interface spectrum \cite{usachov2015}, demonstrating the electronic
decoupling provided by the Au spacer.
However, both Pt-based systems show significantly larger hybridization gaps 
and, as a result, the graphene Dirac cones become substantially distorted.
This effect is especially pronounced in the \grxy{Pt}{Mn}
[Figs.~\ref{fig:orbital}(d1) and \ref{fig:orbital}(d2)] system, where the
nearly flat Mn-derived bands above the DP hybridize only weakly with the
Dirac cone, whereas the dispersive Pt-derived bands below the DP strongly
reconstruct its lower branches around $-0.4$~eV, approximately $0.1$~eV
below the DP. Taken together, these results show that Pt-derived bands play
a more active role than Au-derived bands in reshaping the graphene
$\pi$-bands in the structures considered here.
At the same time, the mostly flat Mn bands around the DP in both Mn-based
systems hybridize more weakly with the graphene $\pi$-bands than their
slightly more dispersive Co counterparts.

\begin{table}[!htbp]
\caption{Properties of the graphene Dirac cones obtained from the
$2\times2$ DFT calculations.  $E_{\mathrm D}$ is the Dirac-point energy relative to the
Fermi level.  $\Lambda_{\overline K}$ and $\Lambda_{\overline K'}$ are the
characteristic minority--majority separations near the corresponding Dirac
points, $\Lambda=E_{\mathrm{min}}-E_{\mathrm{maj}}$.}
\label{table:dft2x2}
\begingroup
\setlength{\tabcolsep}{3.5pt}
\centering
\begin{tabular}{lrrr}
\hline
System $(2\times2)$ & $E_{\mathrm D}$ &
$\Lambda_{\overline K}$ & $\Lambda_{\overline K'}$ \\
 & (meV) & (meV) & (meV) \\
\hline
\grxy{Au}{Co} & $-450$ & 12 & 2 \\
\grxy{Au}{Mn} & $-600$ & 10 & 8 \\
\grxy{Pt}{Co} & $-80$  & 23 & 39 \\
\grxy{Pt}{Mn} & $-300$ & 8  & 17 \\
\hline
\end{tabular}
\endgroup
\end{table}

The spin-resolved carbon projections in
Figs.~\ref{fig:orbital}(a3)--\ref{fig:orbital}(d4) further reveal
distinct spin-dependent gap patterns near the Dirac point.
For both Co-based systems, \grxy{Au}{Co} and \grxy{Pt}{Co},
the minority-spin Dirac cone (blue) shows a clear gap opening in both
$\overline K$ and $\overline K'$ valleys, whereas the majority-spin
cone (red) remains nearly gapless.
Thus, although the detailed dispersions and the magnitudes of the
spin splittings remain valley dependent, the same qualitative
spin-resolved gap pattern is preserved in both valleys for the two
Co-based interfaces.

The Mn-based systems behave differently.
In \grxy{Au}{Mn}
[Figs.~\ref{fig:orbital}(b3) and \ref{fig:orbital}(b4)],
both spin-resolved Dirac cones remain essentially gapless in the two
valleys and are mainly separated by the proximity-induced spin
splitting.
In \grxy{Pt}{Mn}
[Figs.~\ref{fig:orbital}(d3) and \ref{fig:orbital}(d4)],
the minority-spin cone is gapped in both valleys, while the
majority-spin cone remains nearly gapless in the $\overline K$ valley
but develops a finite gap in the $\overline K'$ valley.
The latter system therefore exhibits the clearest valley dependence
of the spin-resolved gap structure among the four commensurate models.

These differences are consistent with the combined action of
exchange, SOC, sublattice asymmetry, and the interface-specific
hybridization discussed above.
Such valley dependence is also known to be enhanced by the restricted
periodic stacking pattern imposed by commensurate
supercells~\cite{slawinska2018,voloshina2018}.
It is further reflected in the calculated spin splittings, which differ
significantly between the valleys for all systems except
\grxy{Au}{Mn}, as summarized in Table~\ref{table:dft2x2}.

\begin{figure*}[!t]
  \centering
  \includegraphics[width=0.84\textwidth]{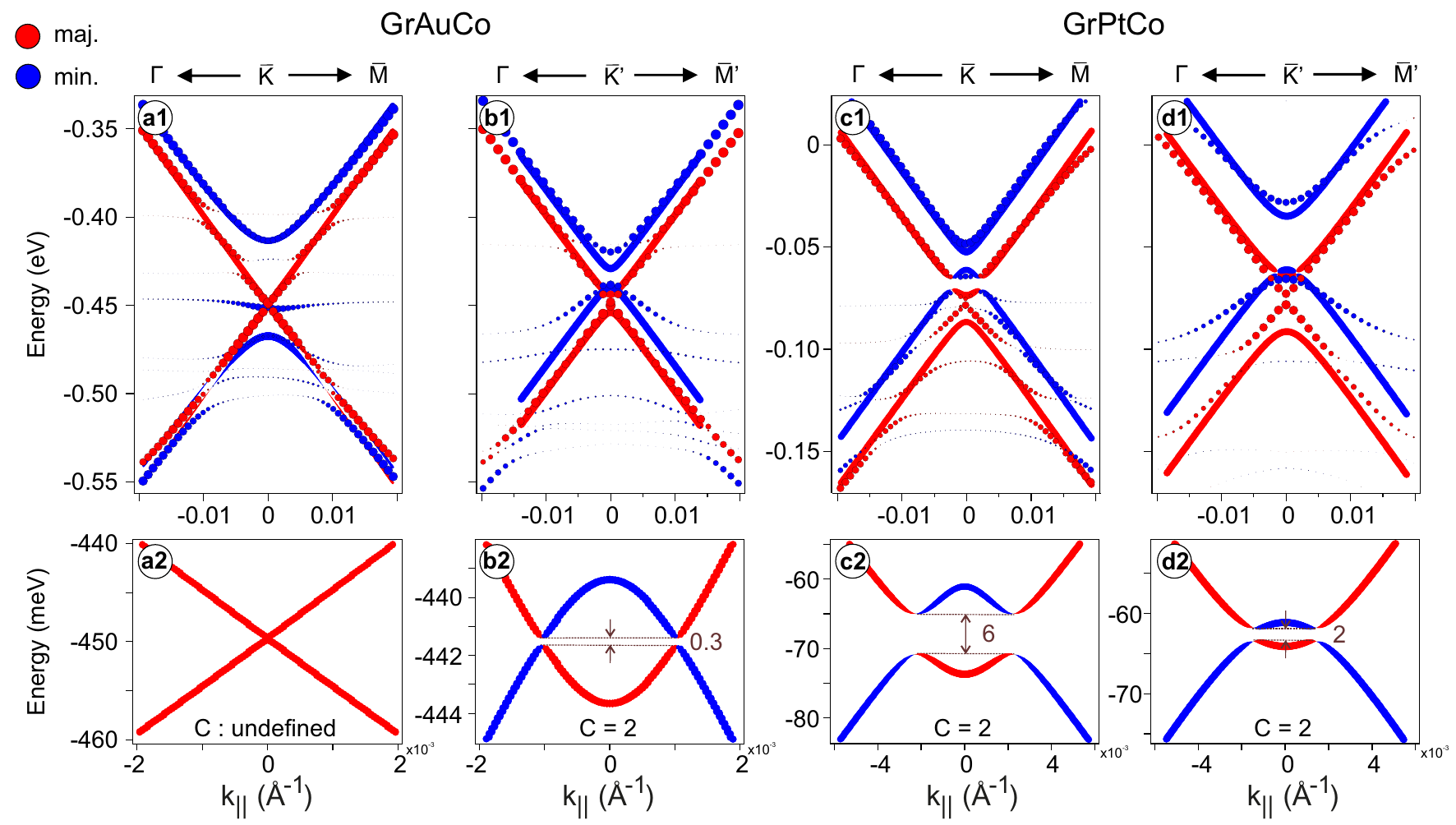}
  \caption{Graphene-projected spin-resolved DFT band structures and
  effective-model fits for the Co-based commensurate $(2\times2)$ interfaces.
  The upper row overlays the fitted bands on the DFT spectra for
  \grxy{Au}{Co} in panels (a1,b1) and \grxy{Pt}{Co} in panels (c1,d1); within
  each pair, the left and right panels correspond to $\overline K$ and
  $\overline K'$, respectively.  Panels (a2)--(d2) show the corresponding
  model bands alone on enlarged energy and momentum scales.  Red (maj.) and
  blue (min.) denote opposite out-of-plane spin polarizations of the carbon
  states; in the upper row, marker size represents their magnitude.  Energies
  in the upper row are relative to the Fermi level.}
  \label{fig:localgap}
\end{figure*}

In order to understand the essential physics of Co-based systems 
revealed by the DFT calculations in commensurate $(2 \times 2)$
models, we have used the tight-binding model to fit the calculated band 
structures (Fig.~\ref{fig:localgap}). 
The strong valley asymmetry enhanced by the periodic stacking pattern of the
commensurate cell
cannot be reproduced by a single set of effective parameters.  We therefore
fit the two valleys separately and use the agreement between the resulting
gapped Hamiltonians as a consistency check of the topological
classification.
With this valley-resolved fitting procedure, both valleys of \grxy{Pt}{Co} yield gapped FM Hamiltonians with pronounced band inversion [Fig.~\ref{fig:localgap}(c,~d)], corresponding to the topological phase with $C=2$ (see Ref.~\cite{eryzhenkov2023}).
The same conclusion (FM case with band inversion and $C = 2$) 
can be drawn only for the $\overline{K}'$ valley in
the case of \grxy{Au}{Co} (Fig.~\ref{fig:localgap}(b)). However,
the fitting procedure for the $\overline{K}$ valley captures only outer 
minority branches of the Dirac cone 
(Fig.~\ref{fig:localgap}(a)) and the inner branches remain gapless, so no
well-defined Chern number is assigned to this fit. 
Therefore, within the applicability of the fitted model,
\grxy{Pt}{Co} is identified with the topological FM phase with $C=2$,
whereas \grxy{Au}{Co} remains topologically indeterminate because its
valley-resolved fits do not provide a common gapped Hamiltonian.

\subsection{Moir\'{e} $9\times9$ structures}

\begin{figure*}[!t]
  \centering
  \includegraphics[width=0.84\textwidth]{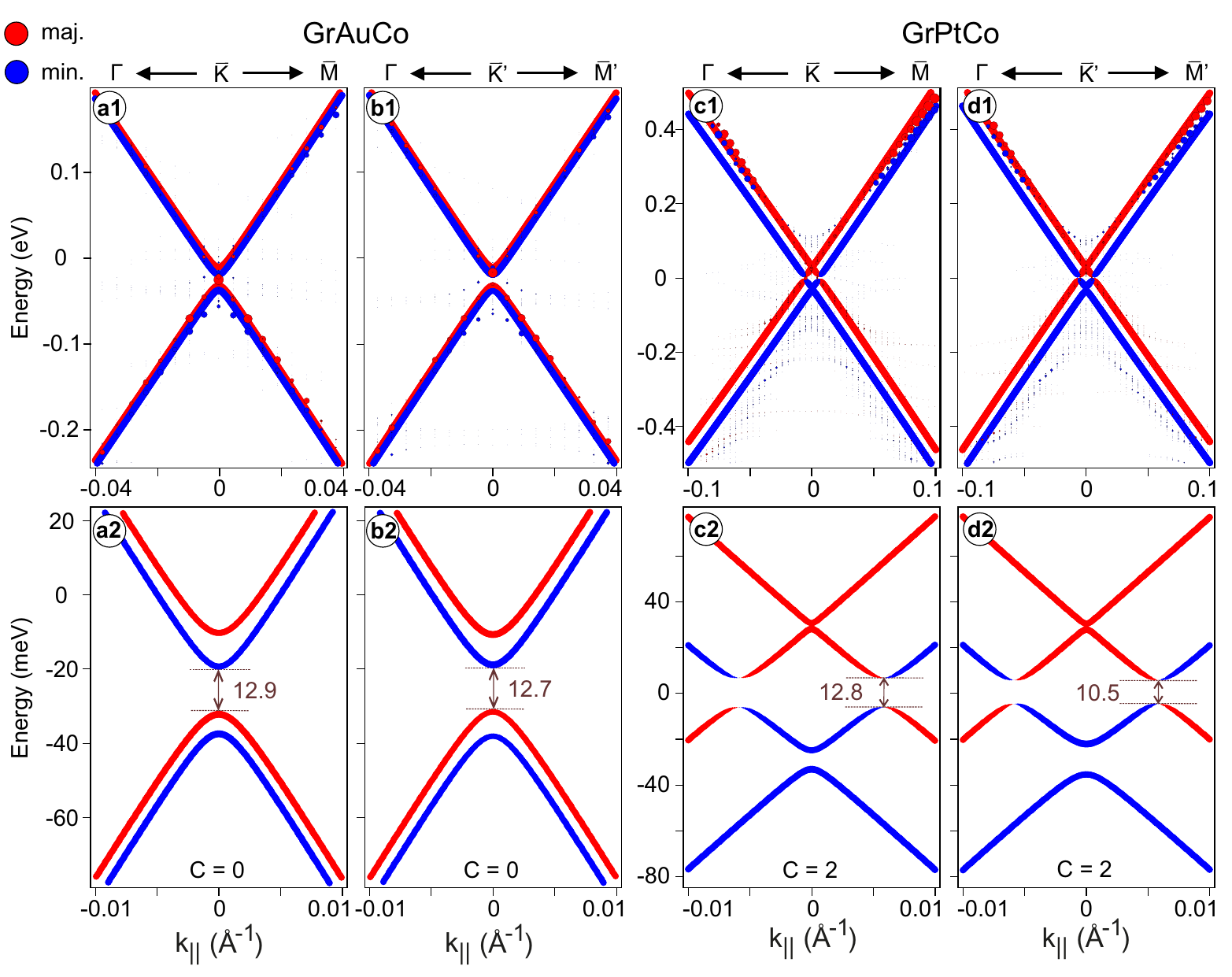}
  \caption{Spin-resolved unfolded carbon-projected DFT band structures
  and effective-model fits for the defect-free moir\'e $(9\times9)$ Co-based
  interfaces.  The upper row overlays the fitted bands on the DFT spectra for
  \grxy{Au}{Co} in panels (a1,b1) and \grxy{Pt}{Co} in panels (c1,d1); within
  each pair, the left and right panels correspond to $\overline K$ and
  $\overline K'$, respectively.  Panels (a2)--(d2) show the corresponding
  model bands alone on enlarged energy and momentum scales.  Red (maj.) and
  blue (min.) denote opposite out-of-plane spin polarizations of the carbon
  states; in the upper row, marker size represents their magnitude.  Energies
  in the upper row are relative to the Fermi level.}
  \label{fig:co9x9}
\end{figure*}

\begin{figure*}[!t]
  \centering
  \includegraphics[width=0.84\textwidth]{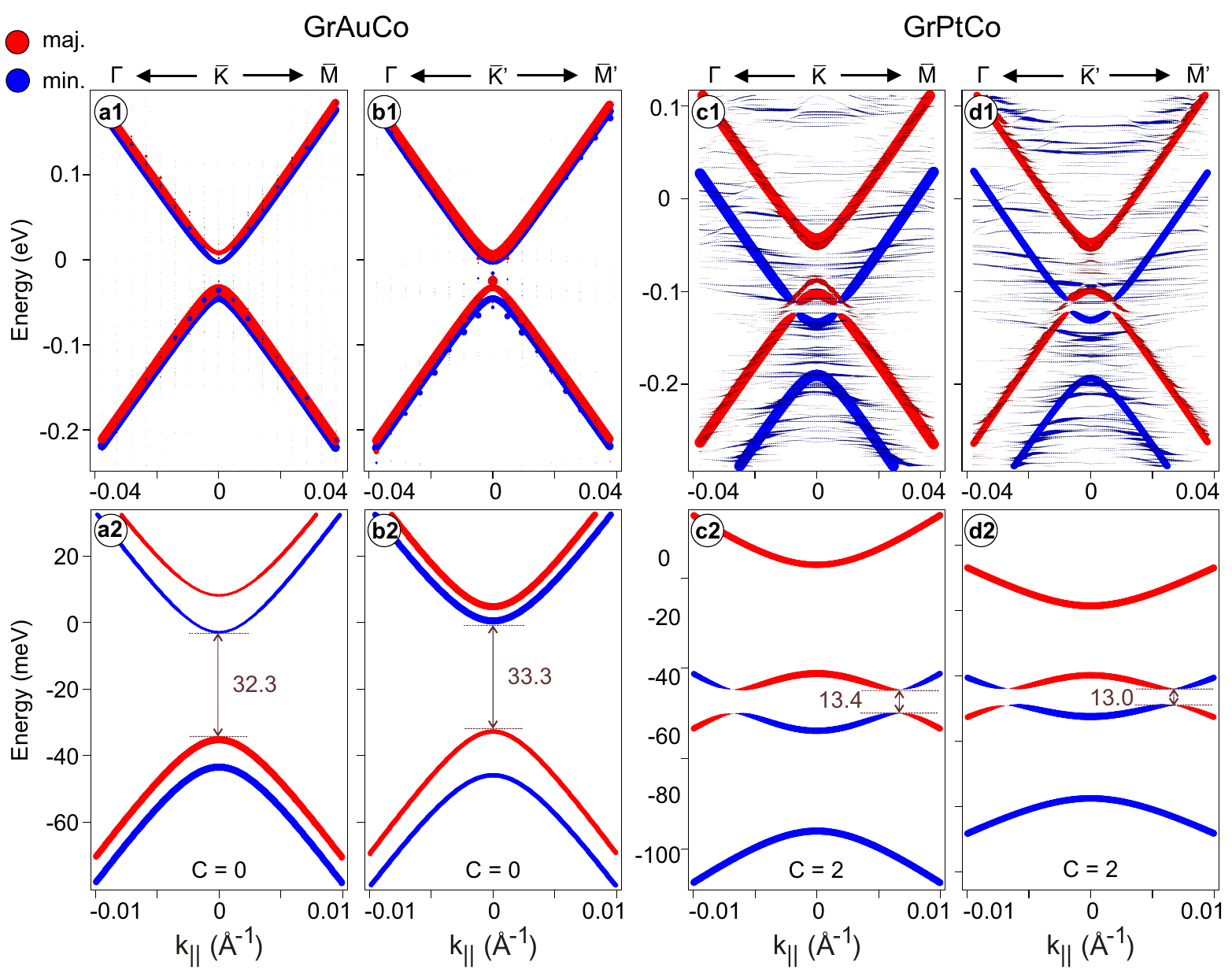}
\caption{Spin-resolved unfolded carbon-projected DFT band structures
and effective-model fits for the moir\'e $(9\times9)$ Co-based interfaces
with a misfit dislocation at the buried noble-metal/Co interface.  The upper
row overlays the fitted bands on the DFT spectra for \grxy{Au}{Co} in panels
(a1,b1) and \grxy{Pt}{Co} in panels (c1,d1); within each pair, the left and
right panels correspond to $\overline K$ and $\overline K'$, respectively.
Panels (a2)--(d2) show the corresponding model bands alone on enlarged
energy and momentum scales.  Red (maj.) and blue (min.) denote opposite
out-of-plane spin polarizations of the carbon states; in the upper row,
marker size represents their magnitude.  Energies in the upper row are
relative to the Fermi level.}
  \label{fig:co9x9disl}
\end{figure*}

\begin{figure*}[!t]
  \centering
  \includegraphics[width=0.86\textwidth]{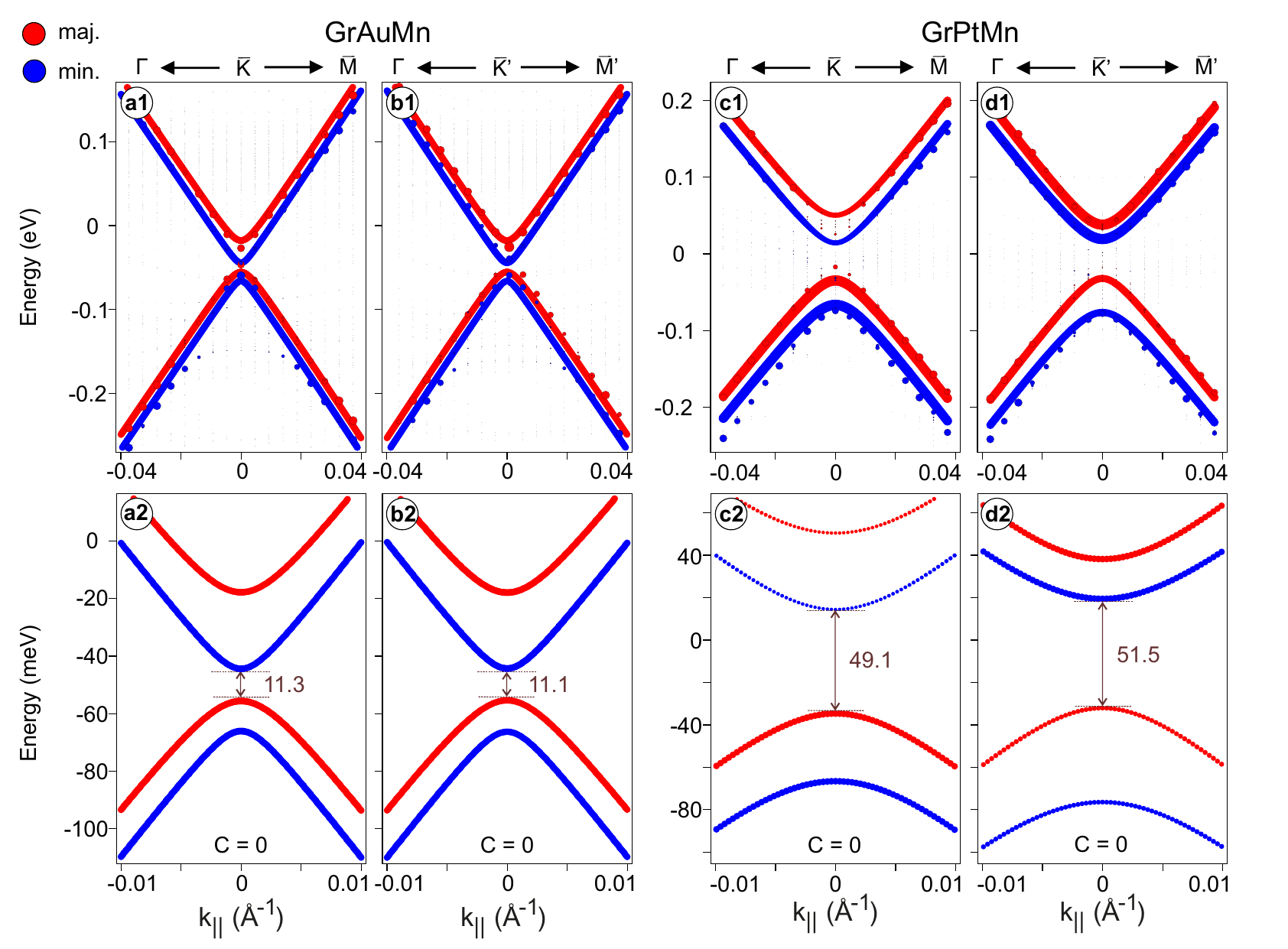}
\caption{Spin-resolved unfolded carbon-projected DFT band structures
and effective-model fits for the defect-free moir\'e $(9\times9)$ Mn-based
interfaces.  The upper row overlays the fitted bands on the DFT spectra for
\grxy{Au}{Mn} in panels (a1,b1) and \grxy{Pt}{Mn} in panels (c1,d1); within
each pair, the left and right panels correspond to $\overline K$ and
$\overline K'$, respectively.  Panels (a2)--(d2) show the corresponding
model bands alone on enlarged energy and momentum scales.  Red (maj.) and
blue (min.) denote opposite out-of-plane spin polarizations of the carbon
states; in the upper row, marker size represents their magnitude.  Energies
in the upper row are relative to the Fermi level.}
  \label{fig:mn9x9}
\end{figure*}

Figures~\ref{fig:co9x9} and \ref{fig:co9x9disl} show the 
unfolded DFT band structures and corresponding tight-binding fits of graphene bands 
of \grxy{Au}{Co} and \grxy{Pt}{Co} for $(9\times9)$ supercells without and 
with the triangular misfit dislocation, respectively. 
Figure~\ref{fig:mn9x9} shows the defect-free $9 \times 9$
supercell computations of the other two systems, \grxy{Au}{Mn} and \grxy{Pt}{Mn}.
Table~\ref{table:dft9x9} summarizes the DP positions and spin splittings
in DFT computations and Table~\ref{table:fit_parameters}
contains the six sets of the corresponding tight-binding model parameters. 

The $9 \times 9$ supercells account for the graphene--metal lattice
mismatch,
so the carbon atoms face a large variety of local stacking registries against
the substrate that are effectively averaged over. In stark contrast to the
$2\times2$ results characterized by significant $n$-doping (hundreds of meV)
of graphene bands and clear valley asymmetry, the $9\times9$ DFT calculations
in all six cases demonstrate an order of magnitude smaller $n$-doping (tens of
meV) and strongly reduced valley asymmetry. The significantly denser
substrate background in the $(9\times9)$ band structures makes the
graphene-derived bands harder to identify because their carbon-projected
spectral weight is fragmented. This makes the tight-binding model particularly
useful for extracting the essential physics governing the graphene-derived
$\pi$-states. The reduced valley contrast allows both valleys to be fitted
simultaneously with one parameter set for every considered structure.

\begin{table}[!htbp]
\caption{Properties of the graphene Dirac cones obtained from the
$9\times9$ DFT calculations.  $E_{\mathrm D}$ is measured relative to the Fermi level, and
$E_g^{\overline K}$ and $E_g^{\overline K'}$ are the direct gaps at the two
valleys.  The characteristic spin splittings are reported as magnitudes,
$\Lambda=|E_{\mathrm{min}}-E_{\mathrm{maj}}|$.  All energies are in meV.}
\label{table:dft9x9}
\begingroup
\setlength{\tabcolsep}{3.0pt}
\centering
\begin{tabular}{lcccc}
\hline
System $(9\times9)$ & Disl. & $E_{\mathrm D}$ &
$E_g^{\overline K}/E_g^{\overline K'}$ &
$\Lambda_{\overline K}/\Lambda_{\overline K'}$ \\
 & & (meV) & (meV) & (meV) \\
\hline
\grxy{Au}{Co} & no  & $-25$ & 13/13 & 12/11 \\
\grxy{Au}{Co} & yes & $-19$ & 32/33 & 12/12 \\
\grxy{Pt}{Co} & no  & $0$   & 13/11 & 57/57 \\
\grxy{Pt}{Co} & yes & $-18$ & 13/13 & 87/89 \\
\grxy{Pt}{Mn} & no  & $-10$ & 49/52 & 29/30 \\
\grxy{Au}{Mn} & no  & $-50$ & 11/11 & 17/17 \\
\hline
\end{tabular}
\endgroup
\end{table}

The reduced valley asymmetry in the moir\'e calculations also makes the
underlying magnetic character of the graphene bands considerably clearer
than in the commensurate models. For \grxy{Au}{Co}, the two valleys now show
closely similar low-energy dispersions
[Figs.~\ref{fig:co9x9}(a1) and \ref{fig:co9x9}(b1)] that are consistently
described by a single FIM model with $C=0$. In this case, the staggered
component of the proximity exchange dominates over the uniform one, producing
the characteristic FIM arrangement of the graphene-derived spin branches.
The corresponding fitted spectra have nearly equal direct gaps in the two
valleys [Figs.~\ref{fig:co9x9}(a2) and \ref{fig:co9x9}(b2)], while the overall
spin splitting remains much smaller than in \grxy{Pt}{Co}.

A qualitatively different pattern is obtained for \grxy{Pt}{Co}.
The band inversion already indicated by the commensurate calculations
persists after moir\'e averaging
[Figs.~\ref{fig:co9x9}(c1) and \ref{fig:co9x9}(d1)], but the much weaker
valley contrast now allows both valleys to be described by the same FM
Hamiltonian with $C=2$. The predominantly uniform exchange separates the
opposite-spin Dirac cones by about $60$~meV and drives their band inversion.
Where branches of opposite spin intersect, the finite Rashba SOC mixes them
and converts the crossings into avoided crossings, thereby opening the gaps
resolved in the fitted bands
[Figs.~\ref{fig:co9x9}(c2) and \ref{fig:co9x9}(d2)].
This exchange-driven inversion followed by SOC-induced gap opening is the
characteristic mechanism of the topological FM sector of the model.
Thus, whereas moir\'e averaging removes much of the artificial valley
asymmetry of the commensurate description, it reveals the essential
contrast between the trivial FIM \grxy{Au}{Co} and the topological FM
\grxy{Pt}{Co} states.

The triangular misfit dislocation alters the band structures of these two Co-based systems
in rather different ways, as reflected in Fig.~\ref{fig:co9x9disl}.
It can be seen clearly that the dislocation strongly enhances the
uniform spin splitting in \grxy{Pt}{Co} from $\approx 60$~meV to
$\approx 90$~meV while leaving its Dirac gaps almost unchanged.
Reconstruction produces complementary changes in \grxy{Au}{Co}: its Dirac gaps
are enhanced from $\approx 10$~meV to $\approx 30$~meV, whereas its spin splittings remain essentially unchanged.
The reconstruction-induced increase of the Dirac gap in \grxy{Au}{Co} to approximately $30$~meV is remarkably consistent with the recent STM/STS study of the same system with a misfit dislocation, which reported a comparable gap under an out-of-plane magnetic field~\cite{rybkin2025nanoscale}.
Although reconstruction substantially reshapes both spectra, their magnetic
and topological classifications are preserved: $C=2$ for \grxy{Pt}{Co} and
$C=0$ for \grxy{Au}{Co}.

The parameter values in Table~\ref{table:fit_parameters} show
that the misfit dislocation slightly decreases the Fermi velocity
($v_\text{F} \propto t$) in both cases but enhances the Rashba coupling
$\lR$ and significantly strengthens the effective exchange interaction.
The latter effect is also spacer-dependent: only the staggered $u_\text{AFM}$
component is enhanced for Au, whereas both exchange components are enhanced for Pt.

Within the Co-based comparison, the Pt-spaced
fit has a larger uniform exchange field and a larger effective Rashba coupling
than the Au-spaced fit.  The comparatively large RMSE of the defect-free
\grxy{Pt}{Co} fit reflects the dense Pt- and Co-derived background, whereas
reconstruction makes the graphene branches more clearly traceable and reduces
the fitting error.

The Mn-based spectra in Fig.~\ref{fig:mn9x9} show that the effective
interactions also depend on the magnetic layer.  \grxy{Pt}{Mn} has a clear
topologically trivial ($C=0$) FIM band structure
with a significant Dirac gap of $\approx 50$~meV.  Table~\ref{table:fit_parameters}
shows that the staggered AFM-like exchange component dominates over the uniform
one in \grxy{Pt}{Mn}. However, the effective Rashba and FM-like exchange couplings in
\grxy{Pt}{Mn} are smaller than in \grxy{Pt}{Co} by factors of approximately
$1.5$ and $2$, respectively.  On the other hand, the topologically trivial
($C=0$) \grxy{Au}{Mn} fit has a stronger FM-like exchange component than
\grxy{Au}{Co}, and graphene therefore exhibits larger spin splittings.
Notably, a carbon-spin constraint was required for \grxy{Au}{Mn}, since the unconstrained carbon moments developed an irregular noncollinear arrangement. By contrast, the \grxy{Pt}{Mn} calculation retained its out-of-plane FIM configuration without a magnetic penalty.

The above analysis establishes the equilibrium topological classification of the considered systems. Their response under photoexcitation, however, is not determined by the Chern number alone. In particular, the topologically trivial ($C=0$) FIM systems retain finite momentum-resolved Berry curvature and can therefore exhibit a Hall response under circular photoexcitation.

\begin{table*}[!t]
\caption{Effective Hamiltonian model parameters directly fitted to reproduce
the graphene bands from all six $(9\times9)$ DFT calculations.  The RMSE
is the root-mean-square band-energy error evaluated jointly over the fitted
points at both valleys.  $E_g^{\mathrm{Gr}}$ is the minimum separation between
bands 2 and 3 over the Brillouin zone and is identified as direct or indirect.
The abbreviation \enquote{Disl.} indicates the presence or absence of the
misfit dislocation in the underlying $(9\times9)$ model.  The FM/FIM
classification is assigned from the relative orientation of the DFT
carbon-sublattice moments: FM denotes parallel moments, whereas FIM denotes
antiparallel moments of unequal magnitude.}
\label{table:fit_parameters}
\centering
\small
\begin{tabular*}{\textwidth}{@{\extracolsep{\fill}}lcccccccccc@{}}
\hline
System & Disl. & Order & $t$ & $\lR$ & $u_\text{FM}$ &
$u_\text{AFM}$ & $\Delta_\text{S}$ & RMSE & $E_g^\text{Gr}$ & $C$ \\
 & & & eV & meV & meV & meV & meV & meV & meV (type) & \\ \hline
\grxy{Au}{Co} & no  & FIM & 3.81 & 2.0 & $-3.4$  & $-10.0$ & $+0.7$ & 6  & 12 (ind.) & 0 \\
\grxy{Au}{Co} & yes & FIM & 3.49 & 5   & $+3.1$  & $-21.0$ & $+1.0$ & 7  & 30 (ind.) & 0 \\ \hline
\grxy{Pt}{Co} & no  & FM  & 3.54 & 12  & $-27.8$ & $-2.7$  & $-3$   & 24 & 11 (dir.) & 2 \\
\grxy{Pt}{Co} & yes & FM  & 3.44 & 18  & $-43.0$ & $-29.0$ & $-0.6$ & 7  & 11 (ind.) & 2 \\ \hline
\grxy{Pt}{Mn} & no  & FIM & 3.43 & 8   & $-12.7$ & $-41.5$ & $-3$   & 7  & 46 (ind.) & 0 \\
\grxy{Au}{Mn} & no  & FIM & 3.70 & 1.6 & $+9.2$  & $-14.9$ & $-4$   & 5  & 11 (ind.) & 0 \\
\hline
\end{tabular*}
\end{table*}

\subsection{Helicity-dependent photoinduced Hall response}
\label{sec:cdhall-results}

\begin{figure*}[!t]
  \centering
  \includegraphics[width=0.96\textwidth]{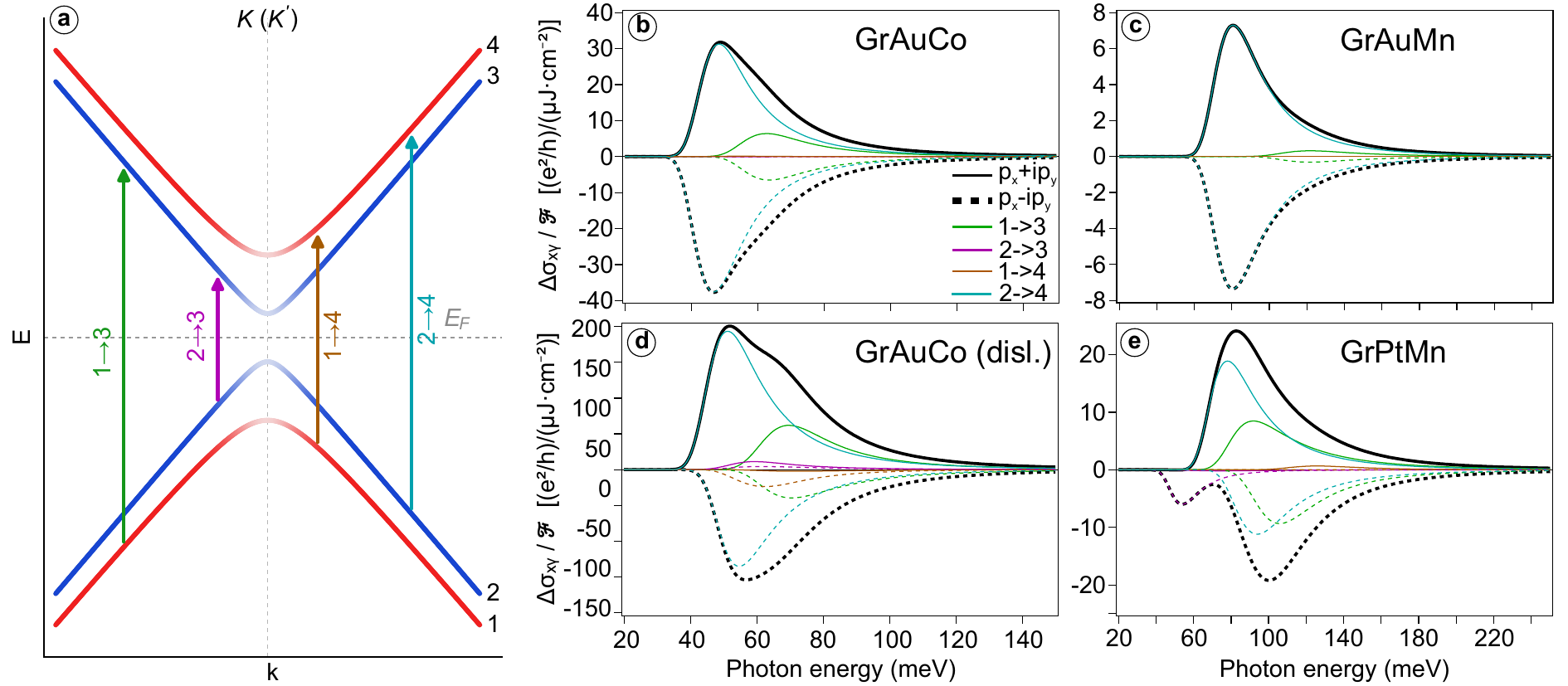}
\caption{Linear fluence-normalized photoinduced instantaneous Hall
response obtained from the fitted FIM $(9\times9)$ models.  Panel (a)
schematically shows the four interband transition channels; red and blue
denote opposite out-of-plane spin polarizations, while the pale segments
indicate spin mixing near the avoided crossings.  Panels (b)--(e) show the
response spectra for the systems identified on the panels; ``disl.'' denotes
the structure containing the misfit dislocation.  Thick black curves give
the total response, and thin colored curves show the contributions of the
individual transition channels.  Solid and dashed curves correspond to
$p_x+ip_y$ and $p_x-ip_y$, respectively.}
  \label{fig:cdhall}
\end{figure*}

We next consider the photoinduced Hall response of the FIM systems. Although their equilibrium Hall conductivity vanishes for $C=0$, circular photoexcitation weights the opposite-valley contributions unequally and thereby generates a helicity-dependent Hall response, as previously demonstrated for quasifreestanding FIM \grxy{Au}{Co}~\cite{rybkin2022}.

Figure~\ref{fig:cdhall} shows the linear fluence-normalized Hall response to circularly polarized infrared excitation in the photon-energy range 20--220~meV. The calculations are based on the defect-free $9\times9$ tight-binding fits for the three FIM systems, with the reconstructed \grxy{Au}{Co} structure included for comparison.

As shown in Fig.~\ref{fig:cdhall}(a), four vertical interband transitions are possible between the two valence bands (No.~1 and~2) and the two conduction bands (No.~3 and~4). The $2\to4$ transition gives the dominant resonance in all four spectra, while $1\to3$ produces a weaker peak or shoulder and $1\to4$ remains weak. The $2\to3$ transition becomes pronounced only at low energies in \grxy{Pt}{Mn}. Since the fitted graphene gaps are indirect, the energies of these vertical optical transitions exceed the corresponding minimum band gaps.

In these FIM band structures, the dominant $2\to4$ and $1\to3$ transitions connect predominantly same-spin branches, whereas the weaker $2\to3$ and $1\to4$ transitions involve predominantly opposite-spin branches. Such dipole \enquote{spin-flip} transitions would be forbidden if spin were a good quantum number, but SOC-induced spin mixing makes them weakly allowed in magneto-spin-orbit graphene. Here, \enquote{spin-flip} therefore refers only to transitions between predominantly opposite-spin branches rather than to transitions between pure spin states. Their strength is governed by the full optical matrix elements and cannot be inferred from the spin character alone. In the defect-free Au-spaced systems, \grxy{Au}{Co} and \grxy{Au}{Mn} [Figs.~\ref{fig:cdhall}(b,~c)], these \enquote{spin-flip} channels are almost completely suppressed. Despite their similar graphene gaps, the two systems exhibit markedly different spectra: \grxy{Au}{Co} shows a broader multichannel response, whereas \grxy{Au}{Mn} is dominated almost entirely by the $2\to4$ resonance.

Compared with defect-free \grxy{Au}{Co} [Fig.~\ref{fig:cdhall}(b)], the
reconstructed system [Fig.~\ref{fig:cdhall}(d)] exhibits a several-fold
larger and broader response.  The low-energy maximum remains dominated by
the $2\to4$ channel, while enhanced $1\to3$ and $2\to3$ contributions extend
the response toward higher photon energies; the $1\to4$ contribution partly
offsets them.  Reconstruction therefore redistributes the response among
the four optical channels as well as increasing its total amplitude.

In \grxy{Pt}{Mn} [Fig.~\ref{fig:cdhall}(e)], the $2\to3$ channel produces
a separate low-energy resonance in the $p_x-ip_y$
channel because $v^{\mathbf k,x} \approx -i v^{\mathbf k,y}$ at the valley
centers. Consequently, the $p_x+ip_y$ matrix element vanishes, whereas the
$p_x-ip_y$ matrix element remains finite. Larger photoexcitation energies
probe regions away from the valley centers, where no such cancellation exists,
so the resulting higher-energy responses are of the same order of magnitude.

Taken together, these results show that the helicity dependence of the photoinduced Hall response is pronounced in all FIM systems except \grxy{Au}{Mn}, although its spectral form and magnitude vary substantially between heterostructures. Misfit-dislocation reconstruction produces the strongest enhancement of the overall response in \grxy{Au}{Co}, whereas \grxy{Pt}{Mn} exhibits the clearest spectral separation between the responses to opposite light helicities.

\section{Discussion}
\label{sec:discussion}

The comparison of the commensurate and moir\'{e} descriptions allows
us to distinguish strongly structure-sensitive spectral properties from
broader topological trends and to test whether the latter survive a realistic
distribution of local stackings. The restricted periodic stacking pattern of
the $(2\times2)$ cells produces excessive charge transfer, pronounced valley
asymmetry, and valley-dependent fits. Moir\'{e} averaging reduces these
effects and permits a common parameter set for both valleys, although the
graphene spectral weight becomes fragmented by the dense metallic background.
Accordingly, a fixed commensurate registry pattern need not yield a
classification that survives moir\'{e} averaging: \grxy{Au}{Co} is
topologically indeterminate in the commensurate model but FIM with $C=0$ in
both moir\'{e} models, whereas \grxy{Pt}{Co} remains compatible with the
topological FM phase with $C=2$ in every representation considered.

Across the moir\'{e} models, the decisive distinction is the character
of the exchange induced on the two graphene sublattices rather than the
magnitude of any single fitted interaction. The two \grxy{Pt}{Co} structures
combine predominantly uniform FM exchange with finite Rashba SOC and have
$C=2$, whereas \grxy{Au}{Co}, \grxy{Pt}{Mn}, and \grxy{Au}{Mn} are FIM and
have $C=0$. In particular, \grxy{Pt}{Mn} retains sizable Rashba coupling and
the largest fitted graphene gap in the series, showing that strong SOC or a
large gap alone does not ensure Chern topology. Within the present
out-of-plane Rashba--exchange model and parameter range, uniform exchange
produces the exchange-inverted band structure and finite SOC gaps the relevant
crossings, whereas staggered exchange and sublattice asymmetry compete with
this mechanism and can stabilize a topologically trivial FIM phase.

The \grxy{Pt}{Co} results further show that a compositionally selected
topological phase can persist across the structural representations considered
here. Reconstruction acts differently in the two Co-based systems: it
selectively enhances the spin splitting in \grxy{Pt}{Co} and the Dirac gap in
\grxy{Au}{Co}, while preserving their effective topological classifications.
The persistence of the $C=2$ assignment in \grxy{Pt}{Co} also gives this
robustness a transport interpretation beyond the binary distinction between
trivial and nontrivial phases. For an isolated gapped graphene-derived sector, $C=2$ corresponds to two co-propagating chiral edge channels and an
ideal Hall conductivity $\sigma_{xy}=2e^2/h$. In the present interfaces,
however, the graphene-derived bands are embedded in a dense background of
metallic substrate states; the fitted $C$ therefore characterizes their
effective topology rather than guaranteeing quantized total transport.
Accessing the corresponding transport regime would additionally require
spectral isolation of the graphene-derived states, suppression of parallel
substrate conduction, and placement of the chemical potential within the
graphene gap.

Related graphene spin-valve experiments demonstrate that
ferromagnetic contacts remain compatible with long-range spin transport when
the interfacial coupling is sufficiently controlled~\cite{Sarkar2025VdWSpinInjection,
Bisswanger2022SpinValves}. These results support the broader feasibility of
spacer-decoupled graphene/ferromagnet architectures and provide an
experimental context for accessing the graphene-derived transport regime
considered here.

The FIM $C=0$ systems nevertheless retain finite valley-resolved
Berry-curvature distributions, which produce distinct helicity-dependent
photoinduced Hall spectra. Reconstruction strongly amplifies the
\grxy{Au}{Co} response, whereas \grxy{Pt}{Mn} provides the clearest
spectrally separated low-energy helicity-selective channel. These differences
provide experimentally distinguishable signatures of the nonequilibrium
response and can be directly probed by helicity-resolved Hall measurements
across the predicted photon-energy windows.

\section{Conclusions}

Our comparison of commensurate, moir\'{e}, and reconstructed
\grall{} interfaces establishes that moir\'{e}-scale registry averaging is
essential for a reliable classification of the graphene-derived bands.  Small
commensurate cells remain useful chemically controlled reference models, but
they can overstate charge transfer and valley asymmetry and produce
valley-resolved topological indications that do not survive moir\'{e}
averaging.  Among the systems studied,
\grxy{Pt}{Co} alone remains in the effective ferromagnetic $C=2$ sector in the
commensurate, defect-free moir\'{e}, and reconstructed descriptions; the
moir\'{e} models of the other compositions are ferrimagnetic with $C=0$.

Across this series, topology is governed chiefly by whether the
proximity exchange is predominantly uniform or staggered on the two graphene
sublattices, rather than by the magnitude of the SOC or the band gap taken
separately.  In the structures considered here, the high-Chern-number sector
is associated with uniform exchange and finite Rashba SOC, whereas staggered
exchange and sublattice asymmetry favor a trivial ferrimagnetic sector.  This
identifies the magnetic order
transferred through the spacer, rather than any single large fitted
interaction, as the central design variable.

Buried-interface reconstruction provides a complementary control.  In
the Co-based systems considered here, it modifies Dirac gaps, spin splittings,
and nonequilibrium Hall spectra without changing the effective topological
classification.  At the same time, the ferrimagnetic $C=0$ systems retain
valley-resolved Berry curvature and a helicity-dependent photoinduced response,
forming a functional class distinct from the effective high-Chern-number
sector.  Composition thus selects the exchange and topological sector,
moir\'{e} modeling determines whether that classification is reliable, and
reconstruction tunes the spectral and optical response.  This separation of
roles provides a compact framework for designing magneto-spin-orbit graphene
interfaces.

\section*{CRediT authorship contribution statement}
\textbf{A.V.~Tarasov:} Conceptualization, Methodology, Validation, Formal
analysis, Investigation, Data curation, Writing -- original draft,
Writing -- review \& editing, Visualization, Software, Supervision, Project
administration, Funding acquisition. \textbf{A.V.~Eryzhenkov:} Methodology,
Validation, Formal analysis, Investigation, Writing -- review \& editing,
Software.

All authors have read and approved the manuscript.

\section*{Funding}
This work was supported by the Russian Science Foundation (Grant No.~25-22-00615).

\section*{Data availability}
The data that support the findings of this study are available from the
corresponding author upon reasonable request.

\end{document}